\documentclass[useAMS]{mn2e}
\usepackage{amsmath}
\usepackage{textcomp}
\usepackage{amssymb}
\usepackage[pdftex]{graphicx}
\usepackage{amssymb}

\usepackage[margin=.8in, paperwidth=8.5in, paperheight=11in]{geometry}
\usepackage{multirow}
\usepackage{multicol}
\usepackage{epsfig}
\usepackage{graphicx}
\begin{document}

\title{Time Dependent Radiation Driven Winds}
\author[S. Dyda, D. Proga]
{\parbox{\textwidth}{Sergei~Dyda\thanks{sdyda@physics.unlv.edu}, Daniel Proga}\\
Department of Physics \& Astronomy, University of Nevada Las Vegas, Las Vegas, NV 89154
}

\date{\today}
\pagerange{\pageref{firstpage}--\pageref{lastpage}}
\pubyear{2018}

\label{firstpage}

\maketitle

\begin{abstract}
We study temporal variability of radiation driven winds using one dimensional, time-dependent simulations and an extension of the classic theory of line driven winds developed by Castor Abbott and Klein. We drive the wind with a sinusoidally varying radiation field and find that   after a relaxation time, determined by the propagation time for waves to move out of the acceleration zone of the wind, the solution settles into a periodic state. Winds driven at frequencies much higher than the dynamical frequency behave like stationary winds with time averaged radiation flux whereas winds driven at much lower frequencies oscillate between the high and low flux stationary states. Most interestingly, we find a resonance frequency near the dynamical frequency which results in velocity being enhanced or suppressed by a factor comparable to the amplitude of the flux variation. Whether the velocity is enhanced or suppressed depends on the relative phase between the radiation and the dynamical variables. These results suggest that a time-varying radiation source can induce density and velocity perturbations in the acceleration zones of line driven winds.

\end{abstract}


\section{Introduction}
Line driving is a viable mechanism for launching winds in massive stars, cataclysmic variables (CVs) and active galactic nuclei (AGN). Simulations have shown that in all but the simplest, spherically symmetric geometries (Castor, Abbott \& Klein 1975, hereafter CAK), line driven winds exhibit complex, time-dependent behaviour. Perturbations at the base of the wind may grow due to the line deshadowing instability  (LDI), leading to a clumpy, non-stationary flow  (Owocki, Castor \& Rybicki 1988, hereafter OCR88; Sundqvist, Owocki \& Puls 2018 hereafter SOP18). In the case of a disc wind, differences in geometry between the flat, equatorial source of photons and matter and the spherically symmetric gravitational potential lead to the growth of density structures at the base of the wind in both 2D axisymmetric simulations (Proga, Stone \& Drew 1998; Proga, Stone \& Drew 1999) and in 3D simulations (Dyda \& Proga 2017 hereafter DP17; Dyda \& Proga 2018, hereafter DP18). Line driving is also sensitive to the ionization state of the gas, which is dependent on the aformentioned density features in the flow, which can act to shield the outermost gas from ionizing radiation (Proga, Stone \& Kallman 2004; Matthews et al. 2016).

In the case of massive stars, one can correlate changes in apparent magnitude, which probe stellar luminosity, to changes in emission lines, which probe the outflows. This has been used to show increases in the luminosity of $\zeta$ Puppis lead to increases of the mass flux of its stellar wind (see Ramiaramanantsoa et al 2018 and refrences therein). In the case of AGN, studies of BAL QSOs have shown that approximately 5\% of systems exhibit disappearing BALs on time-scales of months to years (see De Cicco et al. 2018 and refrences therein). Though such variations are typically explained via changes in covering fraction or ionization state of the absorbers, the dynamics of the absorbers or clouds may be highly sensitive to intrinsic source variability. By comparing the time-variability of spectral lines relative to the continuum using reverberation mapping techniques (for a review see for example Peterson 2001), one can model the distribution and velocity of gas in the system. A key question is how to distinguish between changes in flux due to intrinsic source variations and due to variations in the column density of interposing gas between source and observer. By self consistently modelling the source variation and its effects on the outflow, it may be possible to break the degeneracy of observed flux variations due to intrinsic source variability and due to interposing gas clouds.     

To develop a physical intuition for the effects of temporal source variability on winds, we explore time-dependent models of CAK type 1D spherical line driven winds. This approach is motivated by work by Waters and Proga (2016) that showed that a time varying radiation source can accelerate AGN clouds more efficiently than a stationary source. Other groups have shown that properly treating the line driving may require a more accurate description of the radiation transfer to capture the LDI (SOP18), multiple resonance surfaces in the wind (Springmann 1994) and correctly inferring line opacities from ionization states of the gas (Higginbottom et al. 2014). As a first step to understanding time variability, we use the simplest model for radiation transfer and work in the Sobolev approximation, but vary the driving flux. This will allow us in future work to study time varying models in 2D, and eventually 3D, and compare to our previous work on disc winds, DP17 and DP18, using the Sobolev approximation.    

The CAK solution is a particularly simple case to study because winds are stationary and fully characterized by their physical parameters at the critical point, where the monotonicaly increasing velocity satisfies the relation, $dv/dr|_c = v_c/r_c$, where $v$ and $r$ are the velocity and radial position, respectively and the c subscript denotes the critical point. Thus the solution is characterized by a single physical scale $r_c$ and dynamical time scale $\tau_c = r_c/v_c$. 

For time-dependent winds we expect different solutions depending on the ratio between the dynamical time and the period of the source $T_S$. Driving the wind on long time scales relative to the dynamical time, $T_S \gg \tau_c$, the solution oscillates between the high and low flux stationary states. Driving the wind on short time scales, $T_S \ll \tau_c$, the solution behaves like the mean flux stationary solution. When time scales are comparable $T_S \approx \tau_c$, velocity perturbations with amplitude comparable to variation in driving radiation are induced at the base of the wind. Surprisingly, these velocity perturbations are greater (smaller) than the velocity of stationary solutions, when the phase between the driving radiation and the velocity perturbations are negative (positive). 

In Section \ref{sec:setup}, we describe our simulation setup and our model for intrinsic source variation. In Section \ref{sec:results}, we describe our results for a fiducial run and the dependence of winds on driving source frequency. In Section \ref{sec:discussion} we discuss some implications for observations of massive stars, CVs and AGN. We conclude in Section \ref{sec:conclusion} where we discuss some of the limitations of our approach and possible improvements in future work.  

\section{Simulation Setup}
\label{sec:setup}

Consider a point source of radiation, surrounded by a spherically symmetric, isothermal shell of gas that is optically thin to the continuum radiation. The basic equations for single fluid hydrodynamics are
\begin{subequations}
\begin{equation}
\frac{\partial \rho}{\partial t} + \nabla \cdot \left( \rho \mathbf{v} \right) = 0,
\end{equation}
\begin{equation}
\frac{\partial (\rho \mathbf{v})}{\partial t} + \nabla \cdot \left(\rho \mathbf{vv} + \mathbf{P} \right) = - \rho \nabla \Phi + \rho \mathbf{F}^{\rm{rad}},
\end{equation}
\begin{equation}
\frac{\partial E}{\partial t} + \nabla \cdot \left( (E + P)\mathbf{v} \right) = -\rho \mathbf{v} \cdot \nabla \Phi + \rho \mathbf{v} \cdot \mathbf{F}^{\rm{rad}},
\label{eq:energy}
\end{equation}
\label{eq:hydro}%
\end{subequations}
where $\rho$ is the fluid density, $\mathbf{v}$ the velocity, $\mathbf{P}$ a diagonal tensor with components P the gas pressure, $\Phi = -GM/r$ is the gravitational potential of the central object and $E = 1/2 \rho |\mathbf{v}|^2 + \mathcal{E}$ is the energy where $\mathcal{E} =  P/(\gamma -1)$ is the internal energy. The radiation force is $\mathbf{F}^{\rm{rad}}$ and is described below. We take an equation of state $P = \rho^{\gamma}$ where $\gamma = 1.01$. The isothermal sound speed is $a^2 = P/\rho$ and the adiabatic sound speed is $c_s^2 = \gamma a^2$.  We compute the temperature from the internal energy via $T = (\gamma -1)\mathcal{E}\mu m_{\rm{p}}/\rho k_{\rm{b}}$ where $\mu = 0.6$ is the mean molecular weight, $k_{\rm{b}}$ is the Boltzmann constant and $m_{\rm{p}}$ is the proton mass.

We model the radiation force as a sum of contributions from electron scattering and line driving 
\begin{equation}
\mathbf{F}^{\rm{rad}} = \mathbf{F}^{\rm{rad}}_e + \mathbf{F}^{\rm{rad}}_{LD},
\end{equation}  
where the force due to electron scattering is 
\begin{equation}
\mathbf{F}^{\rm{rad}}_e = \Gamma_* \frac{GM}{r^2} \ \hat{r},
\end{equation}
with $\Gamma_* = L_* \sigma_e/ 4 \pi c G M$ the stellar Eddington parameter, $L_*$ the stellar luminosity and $\sigma_e$ the Thompson cross section. The radiation force due to lines is
\begin{equation}
\mathbf{F}^{\rm{rad}}_{LD} = \Gamma_* M(t) \frac{GM}{r^2} \ \hat{r},
\end{equation}    
where the force multiplier uses the CAK parametrization 
\begin{equation}
M(t) = k t^{-\alpha},
\end{equation}  
with the optical depth paramter 
\begin{equation}
t = \frac{\sigma_e \rho v_{\rm{th}}}{|dv_{r}/dr|},
\label{eq:optical_depth_parameter}
\end{equation} 
and line driving parameters $k = 0.2$ and $\alpha = 0.6$, consistent with a gas thermal velocity $v_{\rm{th}} = 4.2 \times 10^{5} \ \rm{cm/s}$. 

To be able to discuss applications of our results to various objects (from O stars and CVs to AGN), we perform our simulations in dimensionless units and later convert to scalings for the objects of interest. The central object has a mass $M = M_{*}$ and the simulation region extends from $r_* < r < 20 \ r_*$. This dynamical range allows the wind to fully accelerate and reach its terminal velocity. The velocity scale is then $v_* = \sqrt{GM_*/r_*}$ and the time scale $t_* = \sqrt{r_*^3/GM_*}$.  We use a logarithmically spaced grid of $N_r = 128$ points and a scale factor $a_r = 1.08$ that defines the grid spacing recursively via $dr_{n+1} = a_r dr_n$. We choose a hydrodynamic escape parameter $\rm{HEP} = GM_*/r_*c^2_s = 10^{3}$, which ensures that thermal driving is negligible (Stone \& Proga 2009; Dyda et al. 2017). At the inner boundary, we impose outflow boundary conditions on $\mathbf{v}$ and $E$ while keeping the density fixed at $\rho_* = 10^{-10} \ \rm{g/cm^3}$ in the first active zone. This density ensures that we are sufficiently resolving the atmosphere at the base of the wind, but also providing sufficient mass to launch a wind given our choice of Eddington parameter.  At the outer boundary, we impose outflow boundary conditions on $\rho$, $\mathbf{v}$ and $E$. 

We explore cases where the Eddington parameter varies as a function of time, oscillating between a high luminosity state $\Gamma_+ = 0.12$ and a low luminosity state $\Gamma_- = 0.08$. We consider a model where
\begin{equation}
\Gamma_*(t) = \begin{cases}
                                   \Gamma_0 & t \leq t_0 \\
  				\Gamma_0 \left[ 1 + A \sin \left((t - t_0)/T_S \right) \right] & t > t_0
  		\end{cases}
\label{eq:gamma(t)}
\end{equation}
where the amplitude $A = (\Gamma_+ - \Gamma_-)/2 \Gamma_0$ and $t_0$ is chosen so the initial wind reaches a steady state. We explore cases where $1.16 \times 10^{-2} \ t_* \leq T_S \leq 1.16 \times 10^{2} \ t_*$. 

We use the 0, + and - subscripts to refer to the steady state winds with fiducial, high and low luminosity respectively. The dynamics of CAK line driven winds are determined by the mass flux at the critical point (denoted by the subscript c) where $dv/dr\left|_c \right. = v_c/r_c$ (see for example Lamers and Cassineli 1999). Our fiducial solution with $\Gamma_* = \Gamma_0$ allows us to determine the characteristic time-scale of the flow, $\tau_c = r_c/v_c = 1.57 \ r_* \ / 0.92 \ v_*  = 1.71 \ t_*$. The characteristic angular frequency is then $\omega_c = 2 \pi/ \tau_c = 3.67 \ t_*^{-1}$.

\section{Results}
\label{sec:results}
%%%%%%%%%%%%%%%%%%%%%%%%%%%%%%%%%%%
\begin{figure}
                \centering
                \includegraphics[width=0.45\textwidth]{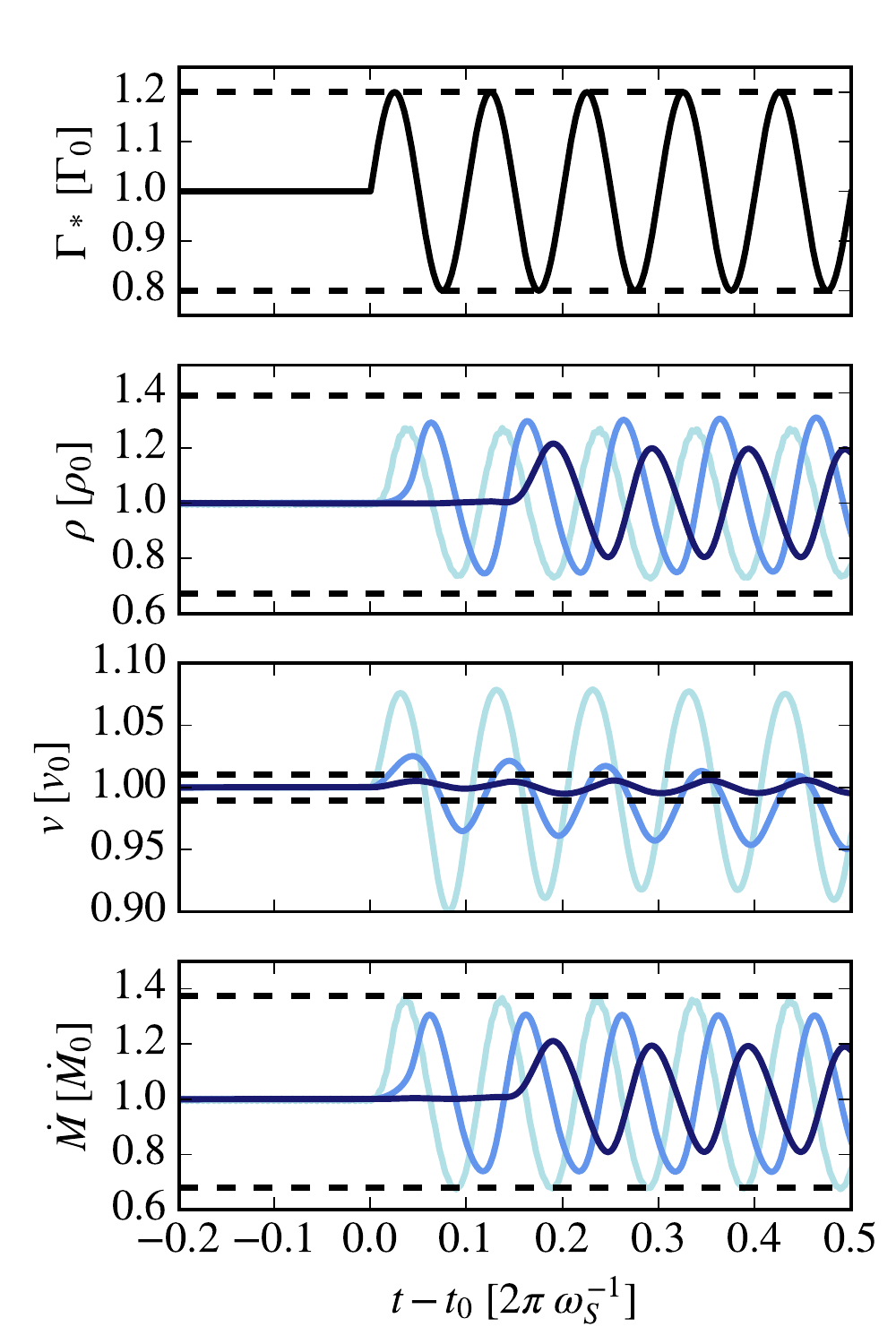}
        \caption{Eddington parameter $\Gamma_*$, density $\rho$, velocity $v$ and mass flux $\dot{M}$ at $r = 2 \ r_*$ (light blue), $r = 5 \ r_*$ (blue) and $r = 20 \ r_*$ (dark blue) as a function of time for $T_S = 0.68 \ \tau_c$. We indicate the values for the high flux $\Gamma_+$ and low flux $\Gamma_-$ stationary states with the horizontal dashed black lines. All plots are normalized to the values of the $\Gamma_0$ stationary solution. Near $t_0$ the solution has not yet reached a periodic state, which is most evident in the velocity profile.}
\label{fig:gamma(t)}
\end{figure} 
%%%%%%%%%%%%%%%%%%%%%%%%%%%%%%%%%%%

%%%%%%%%%%%%%%%%%%%%%%%%%%%%%%%%%%%
\begin{figure*}
                \centering
                \includegraphics[width=0.95\textwidth]{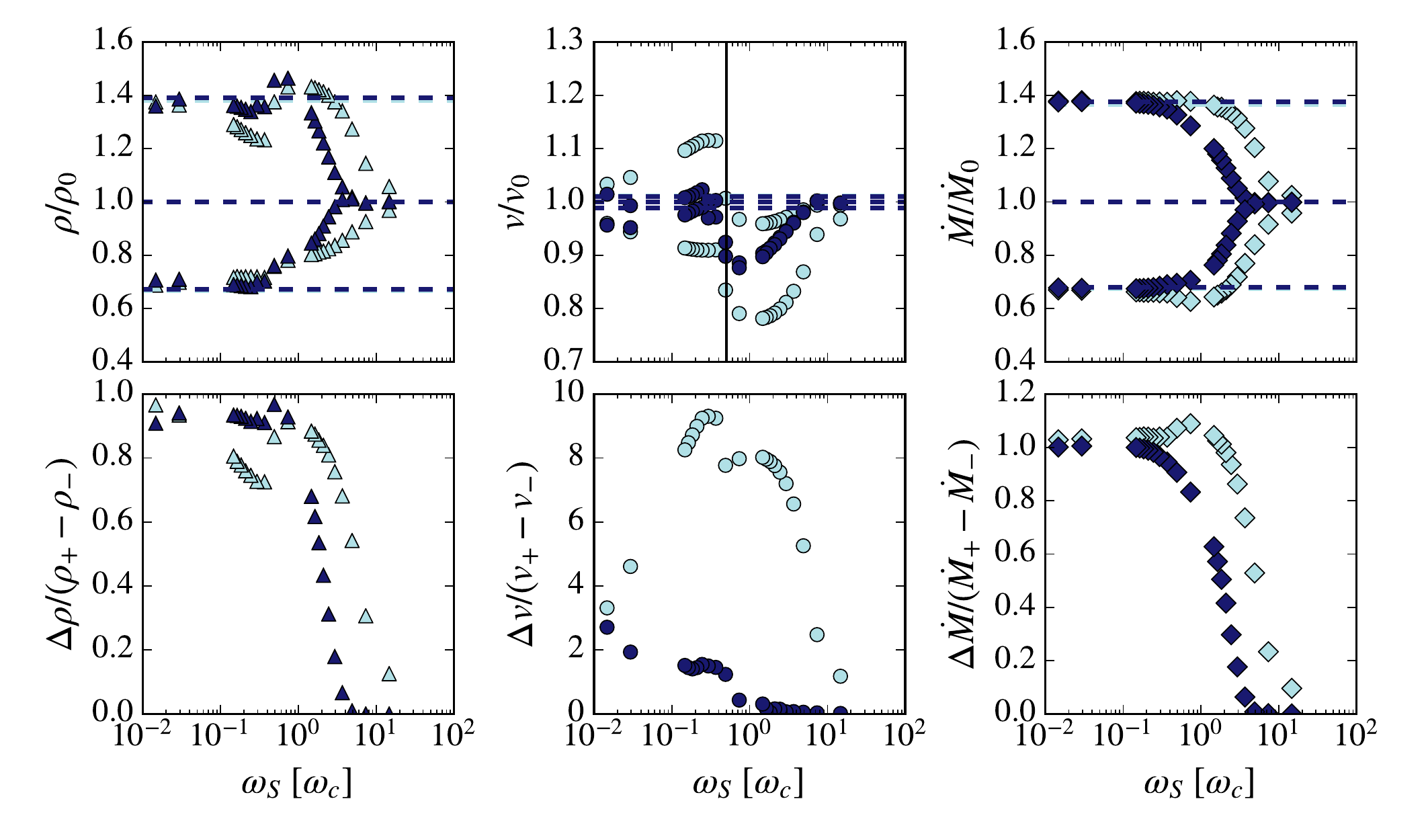}
        \caption{\textit{Upper Panels -} Maximum and minimum values of density $\rho$ (triangles), velocity $v$ (circles) and mass flux $\dot{M}$ (diamonds), normalized to values for the mean radiation solution as a function of driving frequency $\omega_S$ at $r = 2 \ r_*$ (light blue) and $r = 20 \ r_*$ (dark blue). \textit{Lower Panels -} Difference between the maximum and minimum amplitudes for each dynamical variable, normalized to the difference in the high and low luminosity solutions $\Delta X/(X_+ - X_-)$ as a function of driving frequency $\omega_S$. At small angular frequency solutions alternate between the low and high flux solutions. At large angular frequency solutions behave like the mean flux solution. Near the critical frequency we observe a resonance effect, allowing for large enhancements in the velocity.}
\label{fig:frequency_summary}
\end{figure*} 
%%%%%%%%%%%%%%%%%%%%%%%%%%%%%%%%%%%

%%%%%%%%%%%%%%%%%%%%%%%%%%%%%%%%%%%
\begin{figure}
                \centering
                \includegraphics[width=0.45\textwidth]{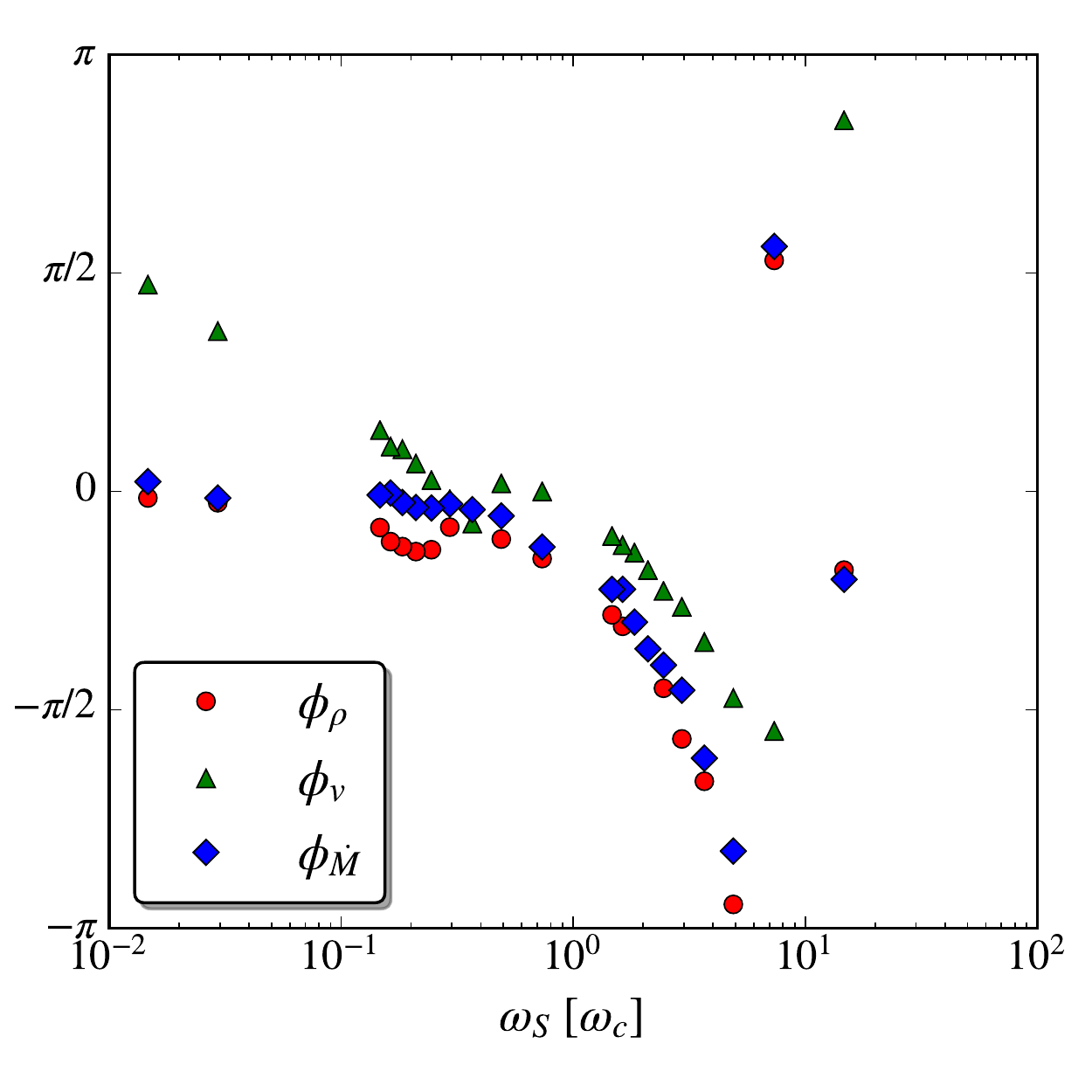}
        \caption{Phase shift $\phi$ between peaks in the Eddignton parameter $\Gamma(t)$ and peaks in the density $\rho$ (red circles), velocity $v$  (green triangles) and mass flux $\dot{M}$ (blue diamonds) as a function of the driving angular frequency $\omega_S$ at $r = 2 \ r_*$.}
\label{fig:phi}
\end{figure} 
%%%%%%%%%%%%%%%%%%%%%%%%%%%%%%%%%%%

We first determined steady state solutions for Eddington parameters $\Gamma_-$, $\Gamma_0$ and $\Gamma_+$ by running simulations for $t \approx 200 \ t_*$. The fiducial run steady state was used as the initial condition for the time varying model (see eq. \ref{eq:gamma(t)}). We find that after a transient period set by the dynamical time, the time-varying solution settles into a quasi-stationary state where the dynamical variables oscillate sinusoidally in time about the fiducial solution
 
\begin{subequations}
\begin{equation}
\rho(t) = \rho_0 + \Delta \rho \sin (\omega t + \phi_{\rho}),
\end{equation}
\begin{equation}
v(t) = v_0 + \Delta v \sin (\omega t + \phi_{v}),
\end{equation}
\begin{equation}
\dot{M}(t) = \dot{M}_0 + \Delta \dot{M} \sin (\omega t + \phi_{\dot{M}}),
\end{equation}
\end{subequations} 
where $\Delta X$ and $\phi_X$ are the amplitude and phase shift of the density, velocity and mass flux. Crucially, both the amplitude and phase shift are functions of the driving frequency, which we further describe below.
 
In Fig \ref{fig:gamma(t)}, we plot the Eddington parameter $\Gamma_*$, the density $\rho$, outflow velocity $v$ and mass flux $\dot{M}$ as functions of time at $r = 2 r_*$ (light blue), $r = 5 r_*$ (blue) and $r = 20 r_*$ (dark blue) for the typical case $T_S = 1.16 t_* = 0.68 \tau_c$. For reference, we plot the high (upper dashed line) and low (lower dashed line) flux solutions. We note the critical point of the fiducial solution, $r_c = 1.57 \ r_*$, so these points are all exterior to the critical point. We plot times near $t_0$ around which the source begins to oscillate.

The transient period lasts approximately for a time $\tau_c$, during which the solution becomes periodic, and dynamical variables begin to oscillate with angular frequency of the source, $\omega = \omega_S$. Oscillations begin at the innermost radii $r = 2 \ r_*$ (light blue) and propagate outwards to $r = 5 \ r_*$ (blue) and to the outermost radius $r = 20 \ r_*$ (dark blue). The phase difference between peaks in the Eddington parameter and in the dynamical variables increases as we go out in radius.  In this case, the amplitude of density oscillations are bounded by the high and low flux stationary solutions, as is the mass flux. On the other hand, the velocity oscillations are not bounded by the high and low flux solutions. At $r = 2 \ r_*$, the amplitude of these velocity oscillations is a factor of a few times the velocity difference in the high and low flux solution. This difference decreases as we move out in radii, with the oscillation amplitude damping with radius.

The magnitude of the oscillations $\Delta \rho, \Delta v, \Delta \dot{M}$ as well as their phase shifts $\phi_{\rho}, \phi_{v}$ and $\phi_{\dot{M}}$ depend non-trivially on $\omega_S$. In the upper panels of Fig. \ref{fig:frequency_summary}, we plot the maximum and minimum values for the density $\rho$ (triangles), velocity $v$ (circles) and mass flux $\dot{M}$ (diamonds), normalized to the steady state values $\rho_0$, $v_0$ and $\dot{M}_0$ respectively for $r = 2 \ r_*$ (light blue) and $r = 20 \ r_*$ (dark blue). The dashed black lines indicate the high and low luminosity steady state solutions. In the lower panels we plot $\Delta X/(X_+ - X_-)$, the difference between the maximum and minimum amplitudes for each dynamical variable, normalized to the difference between the high and low luminosity solutions. 

For small driving frequency $\omega_S \ll \omega_c$, the wind oscillates between the low and high luminosity steady state solutions. This is most apparent in the density and mass flux profiles. At large driving frequencies $\omega_S \gg \omega_c$, the wind resembles the fiducial steady state wind. Near the critical frequency $w_S \sim \omega_c$, we see a smooth transition between these two regimes. Most interestingly, the velocity amplitudes exhibit a resonance effect whereby the difference between the maxima and minima grows beyond the velocity differences between the high/low luminosity states. 

To understand why velocity is enhanced, consider a stationary wind solution, that begins to receive less (more) driving flux. A parcel of gas just beyond the critical point in such a flow will now be more (less) dense than the same parcel of gas in the stationary wind with reduced (enhanced) driving flux. Therefore, it will feel a smaller (larger) acceleration, compared to the reduced flux stationary wind. The mass flux has already been set at the critical point, so this reduced (enhanced) acceleration leads to a slower (faster) wind. Therefore, the wind will be slower (faster) than for a steady accelerating flux. Because we are sinusoidally varying the flux rather than monotonically changing it, we can induce oscillations in the outflow velocity. These fluctuations tend to damp out as you move further out in the wind.  The effects described above are most pronounced in the innermost parts of the flow ($r = 2 \ r_*$) and smaller as we move out in radius to ($r = 20 \ r_*$). At the outermost radius, the source must be varied on timescales roughly 5 times larger to generate a commensurate change in the $\Delta X / (X_+ - X_-)$ profile.  

For driving frequency $\omega \lesssim 1/2 \ \omega_c$, the velocity fluctuations are greater than the fiducial solution, whereas for $\omega \gtrsim 1/2 \ \omega_c$ fluctuations are less than the fiducial solution. This enhancement can be understood by considering the phase difference between the Eddington paramerter and the dynamical variables. In Fig. \ref{fig:phi}, we plot the phase shift for each variable as a function of driving frequency at $r = 2 \ r_*$.  At small driving frequency, density perturbations are in phase ($\phi_{\rho} = 0$) and velocity perturbations are $\phi_v = \pi/2$ out of phase. This is analogous to the case of a simple harmonic oscillator, where the force/acceleration are $\pi/2$ out of phase with the velocity. As the driving frequency approaches $\omega = \omega_c$, the velocity phase difference $\phi_v \approx 0$ before becoming negative at larger driving frequency.

Consider what happens if density is enhanced (suppressed) at the critical point in the stationary flow. Because mass flux is solely determined by the Eddington parameter at the critical point, increasing (decreasing) density at the critical point will decrease (increase) velocity. When $\omega_S \gtrsim 1/2 \omega_c$, both density and velocity lag the driving radiation i.e. $\phi_{\rho}, \phi_{v} < 0$. This means that the positive acceleration phase occurs when the flow is denser, and hence the negative acceleration phase occurs when the flow is less dense. Hence the velocity minimum will be enhanced. When $\omega_S \lesssim 1/2 \omega_c$, the density lags the driving radiation $\phi_{\rho} < 0$ but the velocity leads the radiation $\phi_{v} > 0$. The positive accelertion phase therefore occurs when the flow is less dense and the velocity maximum is enhanced. At $\omega_S \approx 1/2 \ \omega_c$, the velocity is nearly in phase with the radiation, $\phi_v = 0$ and there is no significant velocity enhancement. As with the amplitude of the fluctuations, the turnover in the phase angle plot (Fig. \ref{fig:phi}) is shifted to smaller frequencies by a factor of 5 at the outermost radius.

\section{Discussion}
\label{sec:discussion}

We investigated the effect of varying the amplitude of the luminosity variation with $A = 0.01$ and $0.9$. For $A = 0.01$ our results are qualitatively unchanged. Dynamical variables oscillate sinusoidally at late times and we find changes in the velocity as high as $\Delta v / (v_+ - v_-) \approx 10$. The change in the velocity is approximately given by the amplitude of the luminosity oscillations, $\Delta v \approx \pm 0.5\% \ v_0 \sim A/2 \ v_0$, as with our fiducial case $A = 0.2$. In the case $A = 0.9$ results are qualitatively different. The density and mass flux vary sinusoidally, but the velocity profile behaves like a sawtooth wave. The velocity rise-time for $T_S = 1.16 \ t_*$  is $\Delta t_{\rm{rise}} \approx 1 \ T_S$ whereas $\Delta t_{\rm{fall}} \approx 5.3 \ T_S$. Ulmschneider (1970, 1971) has shown that saw-tooth waves can produce shocks, which will heat the gas. Our isothermal treatment may therefore be insufficient for these larger amplitude cases. The solution reaches a quasi-periodic state, with fluctuations occasionally producing spikes in the density or velocity profile. This case is of less physical interest, as the amplitude is large and thus we do not expect the solution to vary as smoothly as for cases with small amplitude. We do not see a clear trend in boosts to the velocity and only mention this case to confirm that at large amplitude the behaviour of our fiducial case breaks down.

Time variability is a ubiquitous feature of outflowing systems. An important question is determining whether this variability is due to changes in the central object or the result of changes in the wind. Fortunately, the mechanism responsible for driving the flow couples the central object to the wind, allowing for the possible lifting of this degeneracy. In the context of AGN, studies of BAL QSOs have shown that $\sim 5\%$ of systems exhibit \emph{disappearing} X-ray absorption features (De Cicco et al. 2018). BAL variability is seen simultaneously, at multiple velocities and hence at multiple radii in the flow. This suggests the need for a global mechanism for producing BAL variability, and not one that only affects individual clouds. One idea is that variability in the X-ray flux produces global changes in the ionization state of the clouds, which in turn changes the acceleration of clouds due to line driving. We explore this scenario in a limited way in the present work, since the CAK model for line driving depends only on the Eddington fraction from the source. More sophisticated model for line driving may determine the ionization state from the X-ray flux, and the line driving from the UV flux. One may then expect correlations betweeen variability in the X-ray and UV features and therefore require multi-wavelength observation campaigns. The other possible scenario involves clouds in the innermost part of the flow shielding gas in the outermost parts as they orbit at the base of the wind.  Simulations in DP17 and DP18 have shown that clumps naturally form in line driven disc winds due to symmetry breaking of the velocity field. They found clumps were primarily restricted to the base of the wind and therefore disappearing absorption troughs should appear on the orbital time-scale, over which clumps orbit into the line of sight. Intrinsic source variability may provide an additional mechanism for clump generation, which may change the size, physical distribution and formation time-scale of clumps. This may be important in the shielding gas scenario.

Another application in the context of AGN is for reverberation mapping studies. The key variable in the reverberation mapping formalism is the transfer function, which describes the emission line response of the gas as seen by a distant observer due to a delta-function pulse of continuum radiation from the source. The simplest models assume a stationary transfer function i.e. any time variability in the line emission is solely due to variability in the continuum. A more complete model would account for a time dependent distribution of gas, accounted for in a time dependent response function. Such non-stationary transfer functions require knowledge of the radial velocity of the outflow. Since in line driven winds the outflow velocity depends on the radiation source, this could lead to enhanced coupling between the continuum and emission line response. Such coupling may be weaker in the case of thermal winds for instance and absent in the case of magnetic driving. 

Time variability of outflows is is also important in the context of CVs, such as BZ Cam which exhibits a fast and rapidly varying wind (Ringwald and Naylor 1998). CIV absorption line equivalent widths vary in time on $\sim 10 \ \AA$, corresponding to velocity dispersions of $\Delta v \sim 2000 \ \rm{km/s}$, comparable to the outflow velocity (Prinja et al. 2000). DP18 found that velocity dispersion of clumps is $\sim 5\% $ of the outflow velocity. The non-stationarity of line driven disc winds may therefore be insufficient to explain highly variable systems such as BZ Cam. As pointed out by Prinja et al (2000), for compact systems such as CVs the variability time-scale is comparable to the flow time. Therefore variability may be due to the structure of the wind changing, due to changes in the driving luminosity for example, and not because clumps in the wind have evolved. This is different from the model considered in this work, where variability is on time-scales shorter than the flow time and induces perturbations in the wind rather than altering the entire wind.

Other CV systems such as IX Vel and V3885 Sgr have much more stationary winds (Hartley et al 2002). On time-scales shorter than the flow time, they exhibit variability of $v \sim 90 \ \rm{km/s}$ and $v \sim 130 \ \rm{km/s}$ respectively, roughly $10 \%$ of the outflow velocity. However, they show little evidence of any density structure. CV systems thus exhibit different types of temporal variability, which suggests the need to explore different mechanims to break stationarity of wind solutions.    

Another possible application of this model is to spectral variations in O stars. Ramiaramanantsoa et al. (2018) used the BRITE-Constellation's high-precision, time-dependent photometry to study the early O-type supergiant $\zeta$ Puppis. After subtracting out a 1.78 d periodic signal due to the stellar rotation, they found stochastic variation of $\sim 20 \ \rm{mmag}$. They showed that these variations in magnitude, which probes the stellar photosphere, are positively correlated with variations in the He II 4686 \ \AA \ emission line, which probes clumps in the wind. 

A variation of 20 mmag corresponds to fractional changes in intensity $A = \Delta I/I \approx 0.02$. These variations were found to be coherent on time-scales of hours. With a stellar radii $r_* \approx 18 \ R_{\odot}$ and wind terminal velocity $v_{\infty} = 2300 \ \rm{km/s}$, we estimate the dynamical time $\tau_c \approx r_*/v_{\infty} = 1.5 \ \rm{h}$. This is comparable to the time scales of variability explored in our model. Previous study of the He II emission line of this system by Eversberg, Lepine \& Moffat (1998) had found that velocity perturbations are largest within $R \leq 2 R_*$ and tend to decrease at larger radii, consistent with our findings that perturbations are largest near the base of the wind. We stress that direct comparison of our model to observations is overly optimistic, since previous work has shown the need to  properly account for radiative transfer in massive stars. However, it serves as further motivation to study models with intrinsic source variability.  

Additional observational campaigns of O and Wolf-Rayet stars are required to further investigate the relationship between intrinsic stellar variability and variability of accelerating clumps. Given the faintness of such stars and the need for high S/N observations, this is well suited for upcoming Thirty Meter Telescope observations.

\section{Conclusion}
\label{sec:conclusion}

Time variability of emission and absorption features is a fairly ubiquitous feature of outflows from O stars, CVs and AGN. Numerical modeling has shown that when outflows are radiation driven, variability can be induced by instabilities (such as the LDI), geometric effets (as in disc wind models) or by intrinsic variability of the source (as in this work). As a first step, in understanding the effects of intrinsic source variability, we have investigated how a time-varying source alters the otherwise stationary CAK solution.  

Our model makes several simplifying assumptions and is only a first step in understanding time variability of line driven outflows. Within the Sobolev approximation, it may be important to have a more complete treatment of the radiation field. The simplest approach may be to account for optical depth effects, whereby continuum radiation is absorbed as radiation propagates outwards through the wind. At large radii, accounting for optical depth of the over/under densities should be small since they will average out over many oscillations. At small radii, the effect may be more important, as flux will depend on the sign of the density perturbation. Beyond treating optical depth effects, one could perform full radiation transfer and allow for the density perturbations in the wind to couple directly to the radiation field. This may introduce additional time-scales into the problem, beyond the source frequency, and generate additional perturbations in the dynamical variables. Such a treatment may be important for line driven disc winds, relevant to CVs and AGN, where clouds form at the base of the wind. 

Beyond how the radiation field is treated, we must also consider the parametrization of the line driving force itself. The CAK model for line driving is the simplest possible parameterization of the line force. In more sophisticated models of line driving (see for example OCR88) the UV flux determines the line driving but the X-ray flux determines the ionization state, and hence the force multiplier. Using photoionization codes, state of the art line lists may be used to compute the force multiplier from the ionization state and temperature of the gas, rather than relying on approximate powerlaw fits of the force multiplier in terms of the optical depth parameter as in CAK (Dannen, Proga et al. in prep). 

Our model assumes the Sobolev approximation is valid. As such, our non-stationary solution is fundamentally different from models where time-dependence is triggered by the LDI. Owocki and Rybicki (1984) showed that for perturbations generated by the LDI, $\phi_v = \pi/2$ and $\phi_{\rho} = - \pi/2$, with additional corrections due to thermal effects of order 1/HEP. These points do not appear in our phase diagram, which suggests that simulations which resolved the LDI would behave differently and exhibit more complex behaviour. 

This model finds that density and velocity perturbations are induced by a time varying radiation source in a non-rotating, spherically symmetric flow. In spherical geometries rotational effects have been shown to increase mass flux and outflow velocities (Friend \& Abbott
1986; Pauldrach et al. 1986) and recent work by Araya et al. (2018) have studied time-dependent transitions between slow and fast rotating solutions.  In more complex geometries, such as line driven disc winds, complex structure form due to symmetry breaking between the spherically symmetric gravitational potential and the axisymmetry of the matter reservoir and radiation field. DP17 and DP18 characterized the formation of 3D non-axisymmetries at the base of such line driven disc winds, so called clouds, assuming the wind is optically thin to the continuum and isothermal. A time varying radiation field, due to a non-stationary rate of accretion for instance, may lead to an enhancement in the rate of cloud formation. Furthermore, simulations by Waters and Proga (2015) have demonstrated that clouds are sensitive to thermal effects. Clouds may grow due to the thermal instability, and their opacity, which controls the strength of the line driving, depends sensitively on their temperature. Therefore understanding the dynamics at the base of the wind requires proper treatment of the thermodynamics and radiation transfer of the clouds. Ultimately the correct treatment of the clouds at the wind base is necessary to resolve the question of whether variability of lines in the wind is a consequence of variability in the ionization state of the gas or variability in the distribution of clouds along the line of sight, the so called covering factor.

Additional driving mechanisms may also be operating in these systems, which may also affect variability. Dyda et al. (2017) showed that thermal driving is sensitive to the relative intensity of UV and X-ray photons, as with line driving. Therefore we expect thermal and line driving effects to be sensitive to the particular SED  in the system. Further, magnetic driven effects may be important, particularly in AGN systems.

\section*{Acknowledgements}
This work was supported by NASA under ATP grant NNX14AK44G.
%%%%%%%%%%%%%%%%%%%%%%%%%%%%%%%%%%%

\label{lastpage}

\end{document}